\documentclass[aps,prd,amsmath,twocolumn,nofootinbib,floatfix]{revtex4-1}
\usepackage{graphicx}
\usepackage{amssymb,color}
\newcommand{\req}[1]{(\ref{#1})}
\renewcommand{\bar}[1]{\overline{#1}}
\renewcommand{\bar}[1]{\overline{#1}}
\newcommand{\half}{{\frac{1}{2}}}

\newcommand{\ket}[1]{\,\left|\,{#1}\right\rangle}
\newcommand{\mbf}[1]{\mathbf{#1}}

\def\Dslash{\raise.15ex\hbox{/}\kern-.7em D}
\def\Pslash{\raise.15ex\hbox{/}\kern-.7em P}
\def\non{\nonumber}
\def\cb{\color{blue}}

\def\t{\tau}
\newcommand{\la}{\langle}
\newcommand{\ra}{\rangle}

\newcommand{\ben}{\begin{displaymath}}
\newcommand{\een}{\end{displaymath}}
\newcommand{\be}{\begin{equation}}
\newcommand{\ee}{\end{equation}}
\newcommand{\bea}{\begin{eqnarray}}
\newcommand{\eea}{\end{eqnarray}}

\newcommand{\eq}[1]{Eq.~(\ref{#1})}

\newcommand{\bfq}{{\bf q}}      \newcommand{\bfx}{{\bf x}}      \def\bfk{{\bf k}}                                                 
\newcommand{\bfb}{{\bf b}}

\def\xm{x^-}

\def\g{\gamma}\def\r{\rho}\def\a{\alpha}\def\bfB{{\bf B}}\def\d{\delta}\def\bfb{{\bf b}}\def\l{\lambda}\def\th{\theta}\def\tz{{\tilde z}}\def\b{\beta}
\begin{document}

\title{\bf  \hskip10cm NT@UW-19-20\\
The Frame-Independent Spatial Coordinate $\tz$:  Implications for Light-Front Wave Functions, Deep Inelastic Scattering, Light-Front Holography, and Lattice QCD Calculations }

\author{Gerald A. Miller}

\affiliation{ 
Department of Physics,
University of Washington, Seattle, WA 98195-1560, USA}

\author{Stanley J. Brodsky}
\affiliation{SLAC National Accelerator Laboratory, Stanford University, Stanford, CA 94309, USA}
                                                                           
\date{\today}

\begin{abstract}
A general procedure for obtaining frame-independent, three-dimensional light-front coordinate-space wave functions is introduced. The third  spatial coordinate, $\tz$ ,   is the  frame independent coordinate conjugate to the light-front momentum coordinate $x={k^+\over P^+}$ which appears in the momentum-space light-front wave functions underlying generalized parton distributions, structure functions, distribution amplitudes, form factors,  and other hadronic observables.  These causal light-front  coordinate-space wave functions are  used to derive a general  expression for the quark distribution function of hadrons as an integral over  the  frame-independent longitudinal distance (the Ioffe time)  between virtual-photon absorption and emission appearing 
in the forward virtual photon-hadron Compton scattering amplitude.    
Specific examples  using models derived from light-front holographic QCD show that the spatial extent of the proton eigenfunction in the longitudinal direction can have very large extent in $\tz$.

 \end{abstract}  
\maketitle     
\noindent

{\it Introduction}

Much recent effort has been devoted to understanding and measuring the generalized parton distributions \cite{Mueller:1998fv,Ji:1996ek,Radyushkin:1996nd} which encode the fundamental structure of hadrons in terms of the three-dimensional momentum-space  coordinates of their quark and gluon constituents.  Recent lattice calculations of quasi-pdfs evaluate a Fourier transform of a matrix element which depends on the spatial separation 
{    between the points of virtual-photon absorption and emission} that appears in the virtual photon-proton Compton scattering amplitude.  It is therefore of considerable  interest to understand  the spatial longitudinal dependence of the virtual Compton amplitude from a causal, frame-independent perspective.  In this paper we  show that 
the frame-independent eigensolutions of  the QCD light-front Hamiltonian that underly hadronic observables can be expressed in terms of a longitudinal spatial coordinate  $\tz$ that is simply related to the {{}spatial separation between a struck quark and the spectators}.  One thus obtains  a frame-independent fully three-dimensional spatial description of hadron structure which complements analyses
using the usual transverse spatial variables \cite{Soper:1976jc,Burkardt:2002hr,Miller:2007uy,Miller:2010nz}.  

{\it The Light-Front Fock Representation}

The light-front expansion of any hadronic system
is constructed by quantizing quantum chromodynamics
at fixed light-front time  $\tau = t + z/c$~\cite{Dirac:1949cp,Brodsky:1973kr,Brodsky:1979qm,Brodsky:1997de,Brodsky:2000ii}.
The LF time-evolution operator   $P^- = i \frac{d}{d\tau} $ can be derived directly from the QCD Lagrangian.
The light-front Lorentz-invariant Hamiltonian for the composite hadrons 
$H_{LF}^{QCD} = P^-P^+ -{\bf P}^2 $ {    ($P^\pm = P^0 \pm P^z$ and boldface   denotes the two-dimensional transverse vectors)} has 
eigenvalues ${\cal M}^2_h$,  corresponding 
to the mass spectrum of the color-singlet states in QCD~\cite{Brodsky:1997de}.

In principle, the complete set of bound-state and scattering
eigensolutions of $H_{LF}^{QCD}$ can be obtained by solving the light-front
Heisenberg equation
$H_{LF}^{QCD}  \ket{\psi_h} = {\cal M}^2_h \ket{\psi_h},
$
where $\ket{\psi_h}$ is an expansion in multi-particle Fock eigenstates
$\{\ket{n} \}$ of the free light-front
Hamiltonian: 
$ \vert \psi_h \rangle = \sum_n \psi_{n/h} \vert \psi_h \rangle. $ 
The light-front wavefunctions 
$\psi_{n/h}(x_i,\bfk_i,\l_i)$ provide a complete, causal, 
frame independent representation of a hadrons,  relating the quark
and gluon degrees of freedom in each $n$-particle Fock state to the  hadronic eigenstate.

{\it Twist-two operators and the need for a longitudinal spatial coordinate\,}

{{} The quark distributions of a hadron are matrix elements of quark operators at light-like separation~\cite{Collins:1981uw,Ji:1998pc,Collins:2003fm,Tanabashi:2018oca}:
\bea&q(x)= \int {dx^-\over 4\pi}e^{i x P^+x^-}\la P|\bar \psi(-{x^-\over 2})\g^+\psi ({x^-\over2})|P\ra,\nonumber\\
&\bar q(x)=-q(-x) ,
\label{t2}\eea
where the notation $(x^-/2)$ refers to the four vector $ (x^-/2, x^+=0, \bfx=0)$;   the LF helicity and flavor labels, as well as the $Q^2$-dependence, are suppressed. The operator $\g^0\g^+$ that appears in the matrix element in $A^+=0$ gauge serves to project~\cite{Brodsky:1973kr} the dependent field operator $\psi$ onto its independent component $\psi_+$, so that field operators and their adjoints appear.     The variable $x$   ranges between $0$ and $1$.}


\begin{figure}[h]
\includegraphics[width=2.7cm,height=2.253cm]{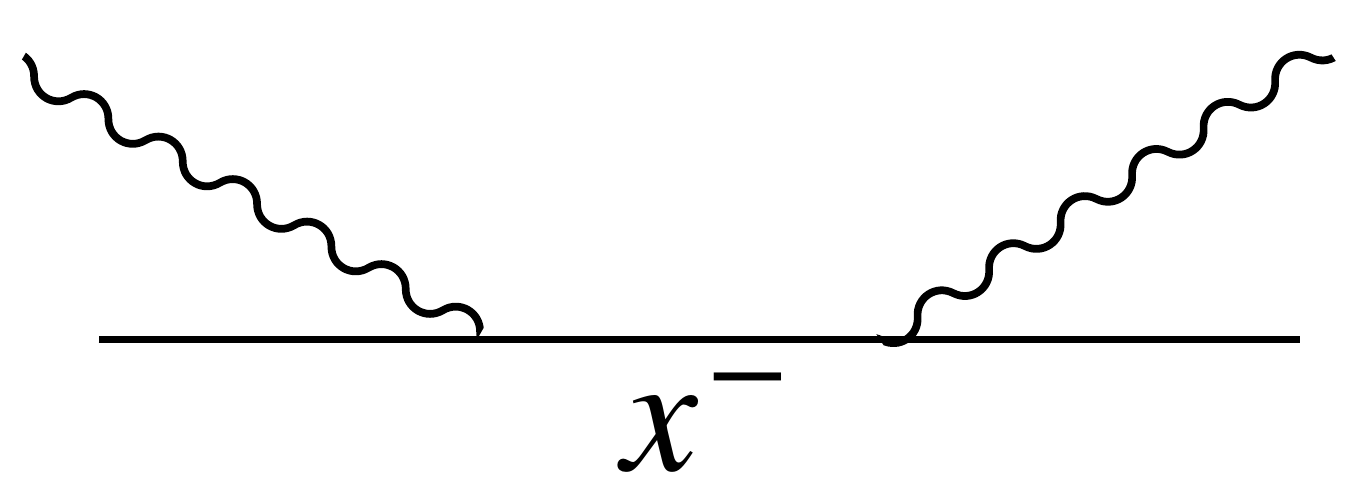}
 \caption{Forward virtual Compton scattering.}\label{VC}\end{figure}
{{} The expression for $q(x)$  }is the leading-twist approximation to  the virtual photon forward scattering amplitude shown in Fig.~\ref{VC}, and $x^-$ is the distance along the light cone between the emission and absorption of the virtual photon. 
{\cb 
}
 The complete  interpretation of the spatial dependence of the quark distributions requires an understanding of their contributions to $q(x)$ as a function of the longitudinal spatial separation $x^-$.

{ 
The  matrix element appearing in \eq{t2} is  directly relevant to   several techniques that seek   to obtain quark distributions as functions of $x$, {\it e. g.} Refs.\cite{Ji:2013dva,Lin:2018qky,Alexandrou:2018pbm,Sufian:2019bol}. See the extensive reviews \cite{Cichy:2018mum,Monahan:2018euv}. 
These techniques   represent  significant advances over efforts based on computing  moments of distributions. Lattice theorists compute {  the lattice version of the } matrix element appearing in \eq{t2}, for example, \cite{Alexandrou:2018pbm},  as $h_{\g^+}(P,x^-),$ and then take a Fourier transform in order to obtain the quasi-pdfs.
 Therefore it is useful  to obtain physical  intuition regarding   the matrix element appearing in \eq{t2}. 
This will be done here by employing recent models  derived from holographic light- front QCD. }

{{}
Of course we are not the first to study the variable $x^-$. It has commonly been called the Ioffe time~\cite{Gribov:1965hf,Ioffe:1969kf,Kovchegov:2012mbw}. This quantity is known to be large if $x$ is small. The study of the {\it matrix element} appearing in \eq{t2}  as the Fourier transform of  quark probability distributions was initiated in Ref.~\cite{Braun:1994jq,Braun:2007wv}.
Our procedure elucidates the dependence on $x^-$ that appears in \eq{t2}  as derived from  light-front wave functions in coordinate space, and  it is thus not  the same as the procedure of Ref.~\cite{Braun:1994jq,Braun:2007wv}.}

{{} 
We
 study}  hadronic light-front wave functions as a function of the longitudinal spatial coordinate of the quark and gluon constituents.  The appearance of wave functions arises by   
inserting a complete set of states $|n-1\ra$ in \eq{t2} so that {{}
\bea& q(x)= 
 2^{1/2} \int {dx^-\over 4\pi}e^{i x P^+x^-}\times\nonumber\\&\sum_n\la P| \psi_+^\dagger(-{x^-\over 2})|n-1\ra\la n-1|\psi_+({x^-\over2})|P\ra. 
\eea} 

The quantity $\la n-1|\psi({x^-\over2})|P\ra$ is an overlap of amplitudes which projects out the active, struck quark, integrated over the spectator particles. This is  simply the light front  Fock space wave function of a quark (or anti-quark). 
In the momentum space representation of the standard Fock space description \cite{Brodsky:1979qm,Brodsky:1997de,Brodsky:2000ii}, one has  for the quark distributions
{{} $\la n-1|\psi_+{({x  
},\bfk, \l})|P\ra\equiv \psi_n(x,\bfk,\l)2^{-1/4},$ in which the indices that refer to specific states have been suppressed to simplify the presentation. 
 The contribution of this component ($q_n$)  $q(x)$  of  \eq{t2} is given by
 \bea q_n(x)= \int {d^2k\over (2\pi)^2}\left \vert\psi_n(x,\bfk)\right\vert^2,\label{fqm}\eea
 {  For quarks $\left \vert\psi_n(x,\bfk,\l))\right\vert^2\propto|\langle n-1|b(k^+,\bfk,\l)|P\rangle|^2$, where $b(k^+,\bfk,\l)$ is the destruction operator and for anti-quarks $\left \vert\psi_n(x,\bfk,\l))\right\vert^2\propto|\langle n-1|d(k^+,\bfk,\l)|P\rangle|^2$,  \cite{Jaffe:1983hp,Diehl:2003ny}. }
}   \\

Converting  these momentum-space wave functions to coordinate space is the next step.   The transverse momentum coordinate $\bfk$ is  transformed into the canonically conjugate impact parameter $\bfb$ to obtain $\psi_n({x},\bfb)$ using standard methods~\cite{Soper:1976jc,Burkardt:2002hr,Miller:2007uy,Miller:2010nz}. 
The dependence {{} of $\psi_n$}  on the frame-independent longitudinal spatial coordinate has not previously appeared.

\smallskip

 {\it The frame-independent longitudinal space coordinate $\tz$}
 
The momentum space wave functions are normally expressed in terms of the longitudinal light-front momentum coordinate $k_i^+\over P^+$, where the index $i$ refers to the $i$'th constituent.  The canonical spatial coordinate is therefore given by the frame-independent  variable  \bea \tz_i=P^+x^-_i.\label{tildez}\eea 
{{} Our $\tz_i$ seems  similar to the variable $z$ of ~\cite{Braun:1994jq}, but its origin and meaning is different. The canonical spatial coordinate occurs for {\it each} of 
the constituents of a Fock space component of a hadronic wave function.   When dealing with the light front wave functions defined above, all of the constituents save one are integrated out. Therefore in the following we use $\tz$ instead of $\tz_i$ to simplify the notation. See also   \cite{Radyushkin:2017cyf,Orginos:2017kos}.}

Making a standard Fourier transform yields
the coordinate space wave function  given by
   \bea \psi_n(\tz,\bfb)={1\over\sqrt{2\pi} }\int_0^1 dx \, \psi_n(x,\bfb)e^{i \tz x},\label{ptz}\eea
   or the mixed form
     \bea \psi_n(\tz,\bfk)={1\over\sqrt{2\pi} }\int_0^1 dx \, \psi_n(x,\bfk)e^{i \tz x}.\label{ktz}\eea
     These light-front (LF) wave functions are independent of the observer's Lorentz frame since both the longitudinal and transverse coordinates are   canonically conjugate to relative LF momentum coordinates.

It is worthwhile to compare the present approach with the concept that the longitudinal direction is Lorentz-contracted to zero in the infinite momentum frame.
Contraction occurs if one  identifies the longitudinal coordinate as $x^-$, the coordinate canonically conjugate to the momentum variable $k^+$. 
This leads to a  frame-dependent  coordinate-space wave function from the relation: 
   \bea  \chi_P^+(x^-,b)=\sqrt{P^+\over{2\pi} }\int_0^1 dx  \psi_n(x,\bfb)e^{i x^-P^+x}\label{c1}\eea
 The resulting density
 $\r_{P^+}(x^-,b)=|\chi_{P^+}(x^-,b)|^2,$ in the infinite momentum frame is obtained by   taking $P^+$ to $\infty$. 
Taking the limit carefully \cite{Dumitru:2018vpr}  yields 
$ \r(x^-,b)=\int {dx }|\psi_n(x,b)|^2\d(x^-),$ 
 a result that  corresponds to a picture in which GPDs are represented as disks \cite{Accardi:2012qut}. 
See also Ref.~\cite{Hoyer:2006xg}.
 There is one caveat: the  contraction occurs {\it only} for   matrix elements of the independent quark-field operators involving the so-called ``good" operator  $\g^+$. \\

However, there is an obvious problem associated with using the frame-dependent $x^-$ coordinate in the infinite momentum frame.  The Lorentz invariant distribution is obtained using {{} instead}
the boost invariant longitudinal coordinate $ \tz=P^+x^-$.
 
 \smallskip
{\it  The variable $\xm $ from light-front wave functions}
\smallskip

  The  frame-independent quark distribution function  $q_n(x)$ of \eq{fqm} can be expressed in terms of the longitudinal coordinate  $\tz$ using the inverse  Fourier transform of \eq{ktz} 
   so that 
   \bea& q_n(x)    =\int {d\tz d\tz'\over 2\pi}\int {d^2\bfk\over (2\pi)^2} \psi_n^*(\tz',\bfk)\psi_n(\tz,\bfk)e^{i(\tz-\tz') x}\non\\&
 \label{F2}  \eea {    Letting  $R\equiv (\tz+\tz')/2$, $x^-= \tz-\tz'$ and  integrating over $R$ yields 
    \bea& q_n(x) =  \int^\infty_{-\infty}  d\xm g_n(x^-,x),\label{gsdef}\eea
 with 
\bea  g_n(x^-,x)=
{1\over 2\pi} \int_0^1 dy\,q_n(y)\cos{x^-(y-x)}.\label{gfinal}
\eea
}

    The function
    $g_n(\xm,x)$  is a measure of the contribution to quark (anti-quark) distribution functions that occur at a particular value of $\xm$. In contrast with the distributions of  ~\cite{Braun:1994jq}, this quantity is real-valued because it is derived from the real-valued quantity of \eq{F2}.

{\it Models to  further our understanding of $g(\xm,x)$ }

  The first model considered is that of a pseudoscalar meson with massless quarks and one valence $q\bar q $ Fock space component.  
  This is the LF holographic model for the massless pion in the chiral $m_q=0$ limit.
 The eigenfunction of the holographic light front Hamiltonian  \cite{Brodsky:2007hb} is given   by: 
  \bea \psi_M(x,\bfb)={\kappa\over \sqrt{\pi} }\sqrt{x(1-x)}e^{-{\bfb^2\kappa^2 x(1-x)\over 2}}.\label{wf}\eea 
The transverse variable~\cite{Brodsky:2007hb} 
 $\zeta^2= b^2 x(1-x)$ is canonically conjugate to $k^2\over x(1-x)$ and the wave functions are simplified if this variable is used. Here we take another path by exhibiting the separate dependence of $\xm$ and  transverse coordinates.

  The  momentum space version of \eq{wf}, relevant for evaluating $g(\xm,x)$, is obtained from the 
 Fourier transform to the canonically conjugate $\bfk$  so that
 \bea \psi_M(x,\bfk)={2\sqrt{\pi}\over \kappa \sqrt{x(1-x)}}e^{-{\bfk^2\over 2\kappa^2x(1-x)}}.\eea
 Using \eq{fqm} one finds that the parton distribution is constant for this model, obtained with massless quarks,
 $ q^M(x)={1 }.$  In contrast, for $m_q\ne 0$ one ~\cite{Brodsky:2014yha}    models the mass-dependence so that  $ q^M(x) = \exp(-{m^2_q\over \kappa^2 x(1-x)})$.

The coordinate space wave function
 is obtained by using  \eq{ptz}. It is useful to    define a light-front coordinate-pace  density:
 \bea \r_M(\tz,b)\equiv |\psi_M^S(\tz,b)|^2. \label{lfd}\eea
 This gives the probability that the struck constituent is separated from the spectators by a longitudinal distance $\tz$  and a transverse separation $b$. Thus obtaining $ \r_M(\tz,b)$ provides a new way of examining hadronic wave functions.
   \begin{figure}[h]
\includegraphics[width=6.6cm,height=5.65cm]{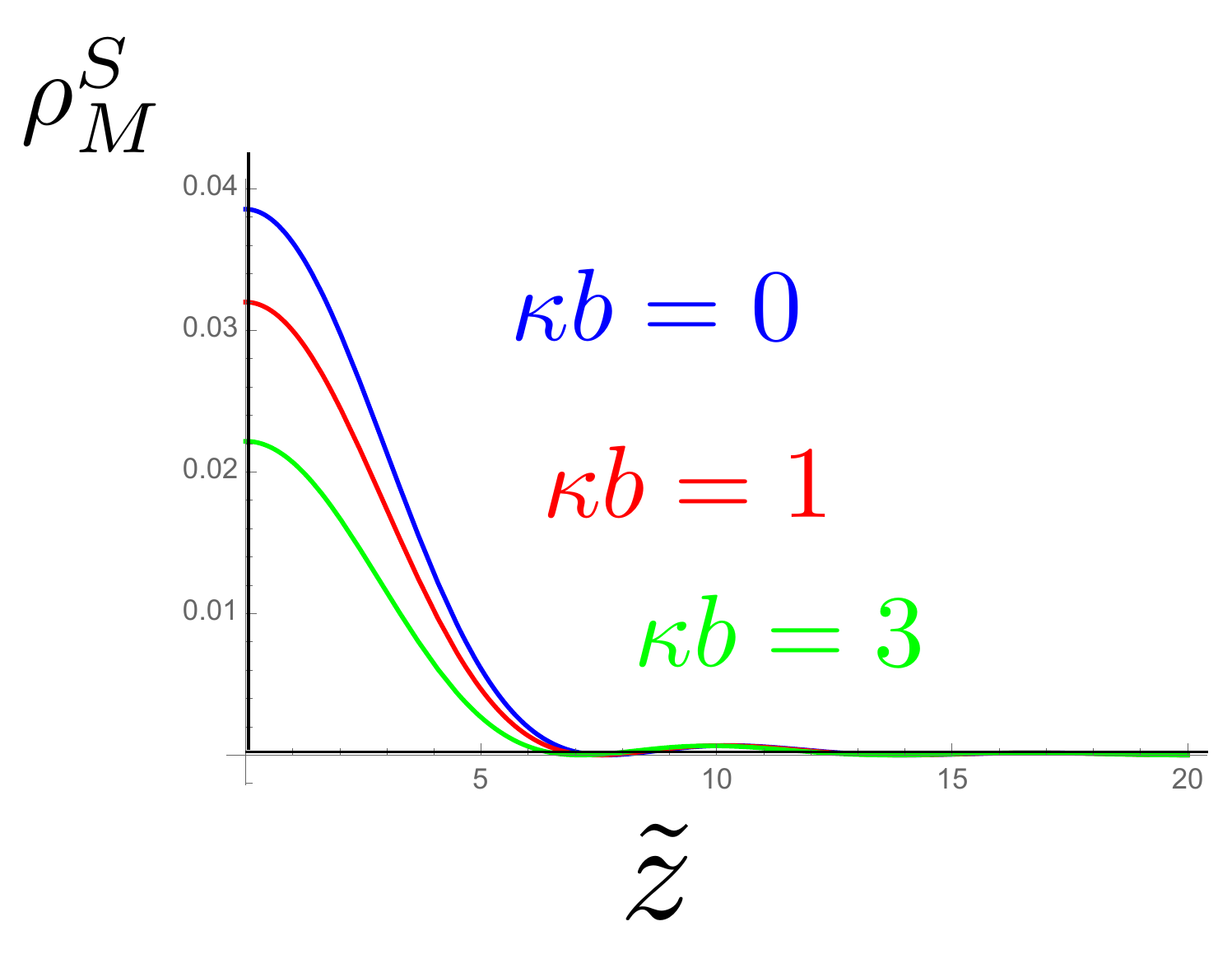}
 \caption{The density $\r^S_M(\tz,b) $ values of $\kappa b=0,1,3 $}\label{fig:diag}\end{figure}
 This   is shown for $m_q=0$  in Fig.~\ref{fig:diag}.    An interesting feature is that
  for each value of $\kappa b$ the density vanishes at  values of $\tz\approx 7.6$. 

  We may gain an understanding of this behavior by obtaining    a closed form expression for $\psi_M^S(\tz,b)$.
By changing variables to $u=x-1/2$ and expanding the exponential in powers of $\l\equiv \kappa^2b^2/8$  we find  the most important term to be 
    \bea  {\sqrt{2\pi}\over 2\kappa}\ \psi_M^S(\tz,b)\approx {\pi\over4}e^{i \tz/2}\,e^{-b^2\kappa^2/8}{J_1(\tz/2)\over \tz},\label{result}\eea
  which is reasonably    accurate for $\tz$ greater than about $b^2\kappa^2$.
   The first zero of  $J_1(x) $ occurs at $x=3.8171$;  one can obtain a qualitative understanding  of this  zero crossing as shown in Fig.~\ref{fig:diag}.
 The result \eq{result} means that for such values, the density falls only as $1/\tz^3$ approximately modulated by  $\cos^2({\tz +\pi/4})$. The existence of such a large-distance tail indicates that using the $\tz$ variable has the potential to reveal interesting aspects of hadronic  physics.
 \\

{ 
The next step is to determine the function $g(\xm,x)$ 
    for the model of \eq{wf}.
    Use \eq{gfinal} with $q_n(y)=1 $ as given above, so that 
          \bea  g_M(\xm,x)= {1\over 2\pi}\frac{\sin (\xm x)+\sin (\xm(1- x))}{\xm}.\eea
 Observe the  slow, $1/\xm$, falloff with increasing values of $\xm$ for all values of $x$. 
 
 Similarly, a     slow falloff is also obtained for models with massive quarks. In the limit of large quark masses, defined by   $\g\equiv m_q^2/\kappa^2>1$, we find that 
     \bea& g_M(\xm,x)
   \approx \frac{e^{-4 \gamma } e^{-\frac{{(\xm)}^2}{64 \gamma }} \cos \left(\xm
   \left(x-\frac{1}{2}\right)\right)}{8 \sqrt{\pi } \sqrt{\gamma }}.
     \eea }
     

    ~~
    
    {\it Universal Light Front Wave Functions \cite{deTeramond:2018ecg}}
    
    ~~
The model given in \eq{wf} is very simple, with  $q_n(x)=1$  for $m_q=0.$ 
 A recent paper  \cite{deTeramond:2018ecg} presents a universal description of generalized parton distributions obtained from Light-Front Holographic QCD, and we  shall use their models for  light-front wave functions. These are presented  as  functions  of the number  $\t$ of constituents of a   Fock space component.   Regge behavior at small $x$ and inclusive counting rules as $x\rightarrow1$ are incorporated. Nucleon and pion   valence quark distribution functions  are obtained in precise agreement with global fits.  {\color{blue}The} model is described by the  
quark distribution  $q_\tau(x)$ and the profile function $f(x)$ with
\bea \label{qx} &
q_\tau(x) = \frac{1}{N_\tau} \big(1- w(x)\big)^{\tau-2}\, w(x)^{- \half}\, w'(x), \\&
 \label{fax}
f(x) =   \frac{1}{4 \lambda}\left[  (1-x) \log\left(\frac{1}{x}\right) + a (1 - x)^2 \right],
 \eea
and $w(x) = x^{1-x} e^{-a (1-x)^2}.$

The value of the universal scale $\lambda$ is fixed from the $\rho$ mass: $\sqrt{\lambda} = \kappa = m_\rho/ \sqrt{2} = 0.548$ GeV~\cite{Brodsky:2014yha,Brodsky:2016yod}. The   flavor-independent parameter  $a = 0.531 \pm 0.037$.   The $u$ and $d$ quark distributions of the proton  are given by  a linear superposition of $q_3$ and $q_4$ while those of the pion are obtained ~\cite{deTeramond:2010ez} from $q_2$ and $q_4$.\\

{ 
  
  \begin{figure}[h]
\includegraphics[width=4.9cm,height=4.9cm]{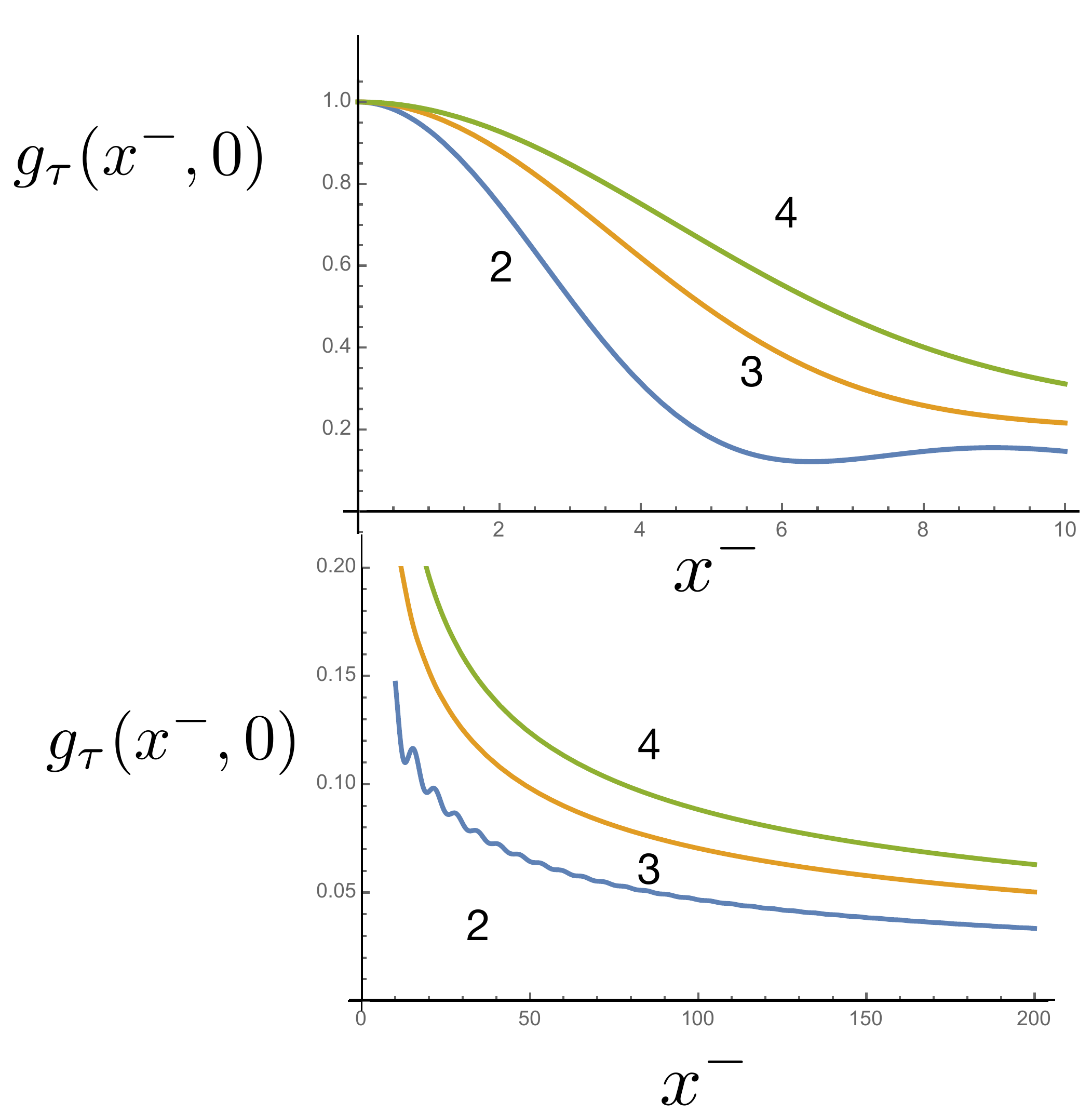}
 \caption{$ g_\t(\xm,0)$. The numbers refer to the value of $\t$, the number of constituents in the Fock state.}\label{gtilde}\end{figure}
 
 Given these distributions we may study the function $g(\xm,x)$ of  as a function of $\t$, using   \eq{gfinal} with $q_\t$ replacing $q_n$.
 Fig.~\ref{gtilde} shows $g(\xm,0)$ as a function of $\xm$, }
the dimensionless separation between the emission and absorption of the photon $\xm$ of Fig.~\ref{VC}.  In the lab frame $x^-= \sqrt{2}/M \xm= 0.3 \xm $~fm, so that the proton radius corresponds to about $\xm=3.$
One observes a  slow falloff with increasing $\xm$: $ g_\t(\xm)\sim {1\over \sqrt{\xm}}$ for all values of $\t$. 
This qualitative behavior can be understood analytically. The function $q_\t(x) \sim1/\sqrt{x}$ for small values of $x$ and $(1-x)^{2\t-3}$. A useful  approximation for  $q_\t(x) $ is given by the product of the two forms. In that case
one may consider {\it e.g.}  
\bea 
&\int_0^1{dy\over\sqrt{y}}\,(1-y)^3\cos{\xm (y-x)}\\&\sim \frac{\sqrt{2 \pi } {(\xm)}^{3/2} \sin (x)+\sqrt{2 \pi } {(\xm)}^{3/2} \cos (x)+3
   \cos (x-{\xm})}{2 {(\xm)}^2}\eea
which,  for all values of $x$,  demonstrates  $(\xm)^{-1/2}$ falloff, modulated by oscillatory behavior. { An  essential  feature is 
that  there is  a significant probability that the deep inelastic scattering process occurs at large separations between the absorption and emission of the virtual photon. }
 
The traditional idea that large longitudinal distances (the Ioffe time)~\cite{Ioffe:1969kf,Braun:1994jq}, underlies  deep inelastic scattering at small $x_{bj}$  is related  to the hadronic light-front wave function. 
 
\begin{figure}[h]
\includegraphics[width=6.198cm,height=5.16cm]{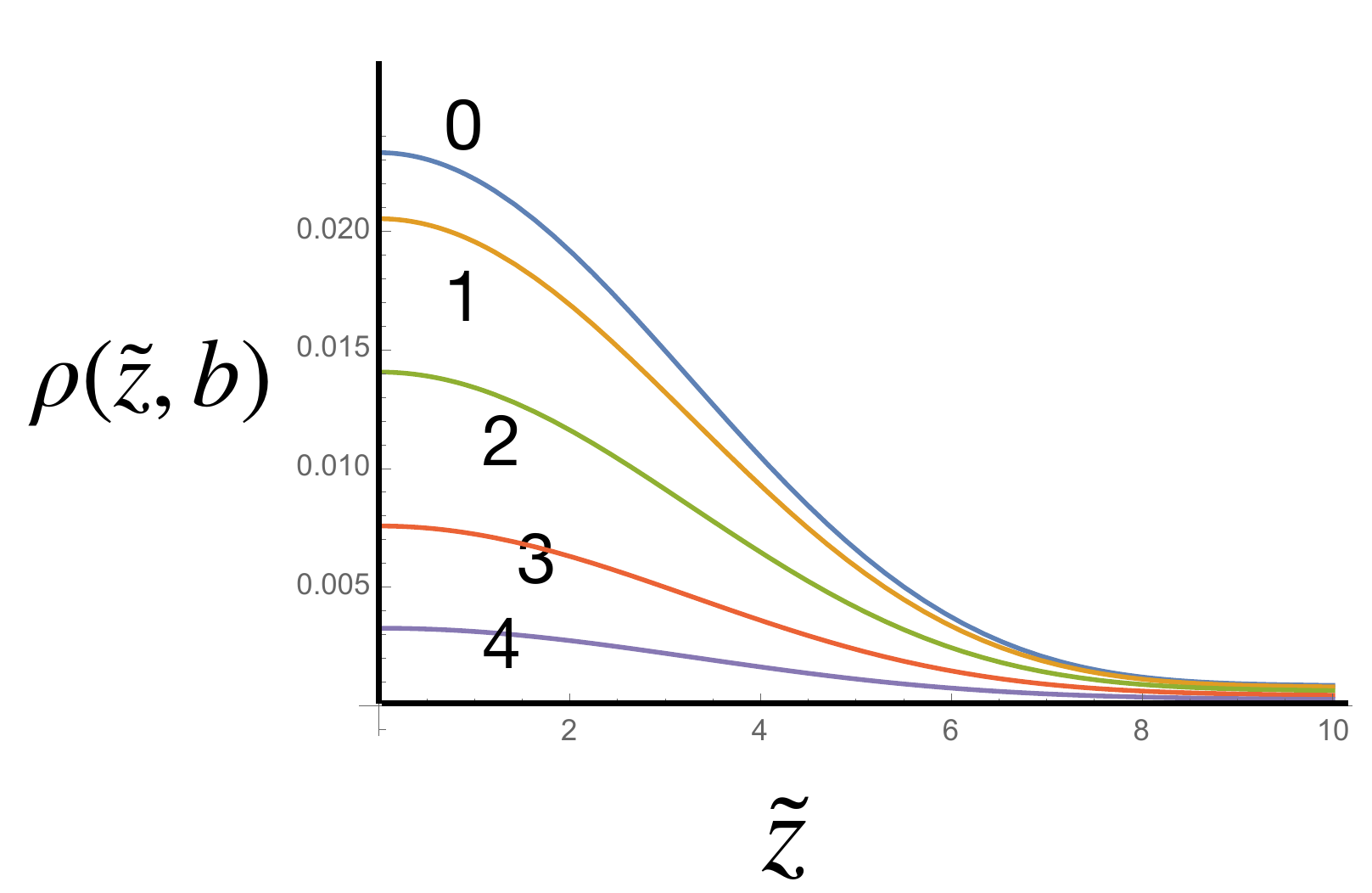}
 \caption{$\r(\tilde z,b)$. The numbers refer to the value of $b$ in units of GeV$^{-1}$.}\label{denzu}\end{figure}

Ref.~\cite{deTeramond:2018ecg} also presents
the universal light front wave function (LFWF): 
\be \label{LFWFb}
\psi_{\rm eff}^\tau(x, \mbf{b} ) = \frac{1}{2 \sqrt{\pi}} \sqrt{\frac{q_\tau(x)}{f(x)}} 
 (1-x) \exp \left[ - \frac{(1-x)^2 }{8 f(x) } \, \mbf{b}^2\right],
\ee
in the transverse impact space representation with $q_\tau(x)$ and $f(x)$ given by \req{qx} and \req{fax}. The   dependence on $\tz$, is contained in the wave function  $\psi_\t(\tz,b)$, computed  according to \eq{ptz}.  The density $\r(\tz,b)=\left\vert\psi_{\rm eff}^\tau(\tz, \mbf{b} ) \right\vert^2$ is shown in Fig.~\ref{denzu}.
 \begin{figure}[h]
\includegraphics[width=5.398cm,height=3.6cm]{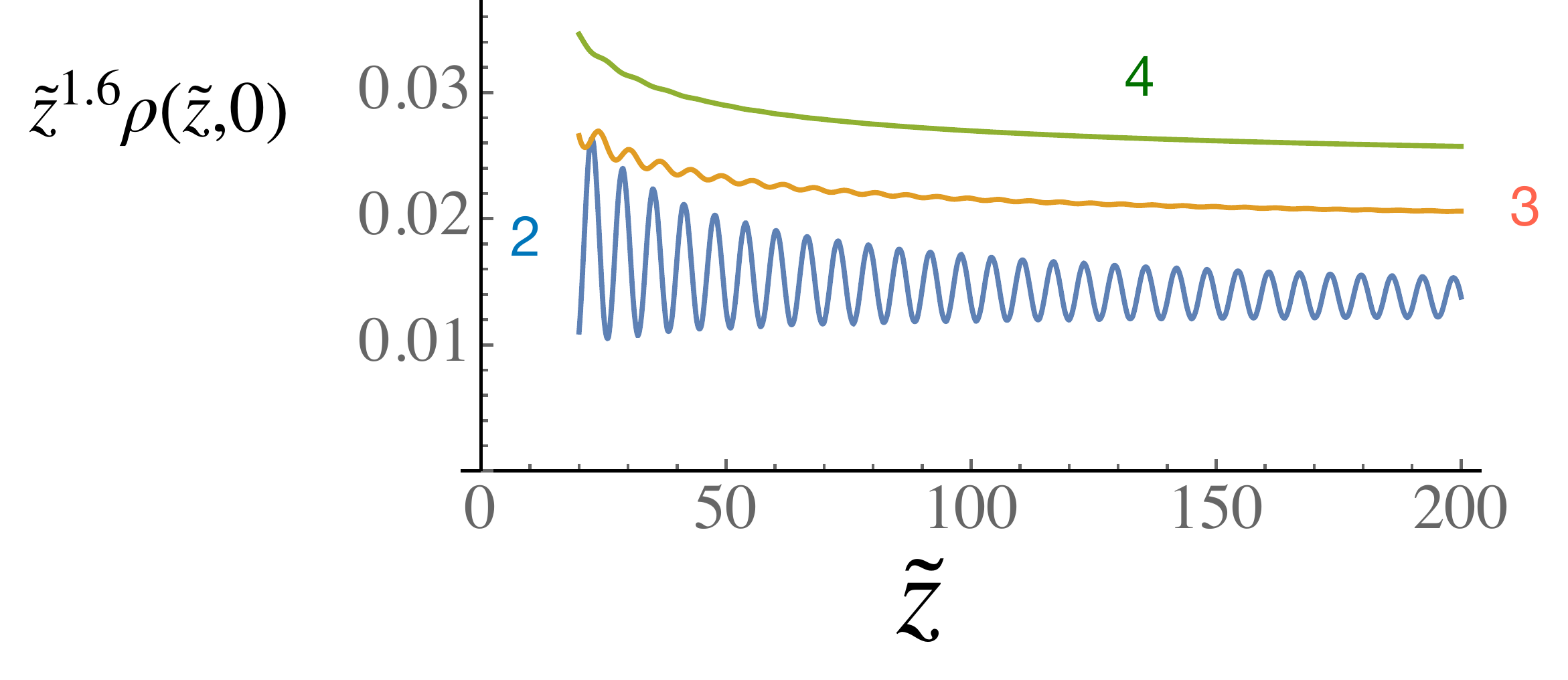}
 \caption{$\tz^{1.6}\r(\tz,0)$. The numbers refer to the value of $\t$,  the number of constituents in the Fock state.}\label{zdep}\end{figure}
 
The main interest here is to study the dependence on $\tz$, which is displayed in Fig.~\ref{zdep} for $b=0$. The same general behavior is seen for other values of $b$. The density falls roughly as $1/\tz^{1.6}$. This very slow falloff that again indicates the  large spatial  extent of hadronic wave functions.
 \\

\begin{figure}[h]
\includegraphics[width=5.398cm,height=4.6cm]{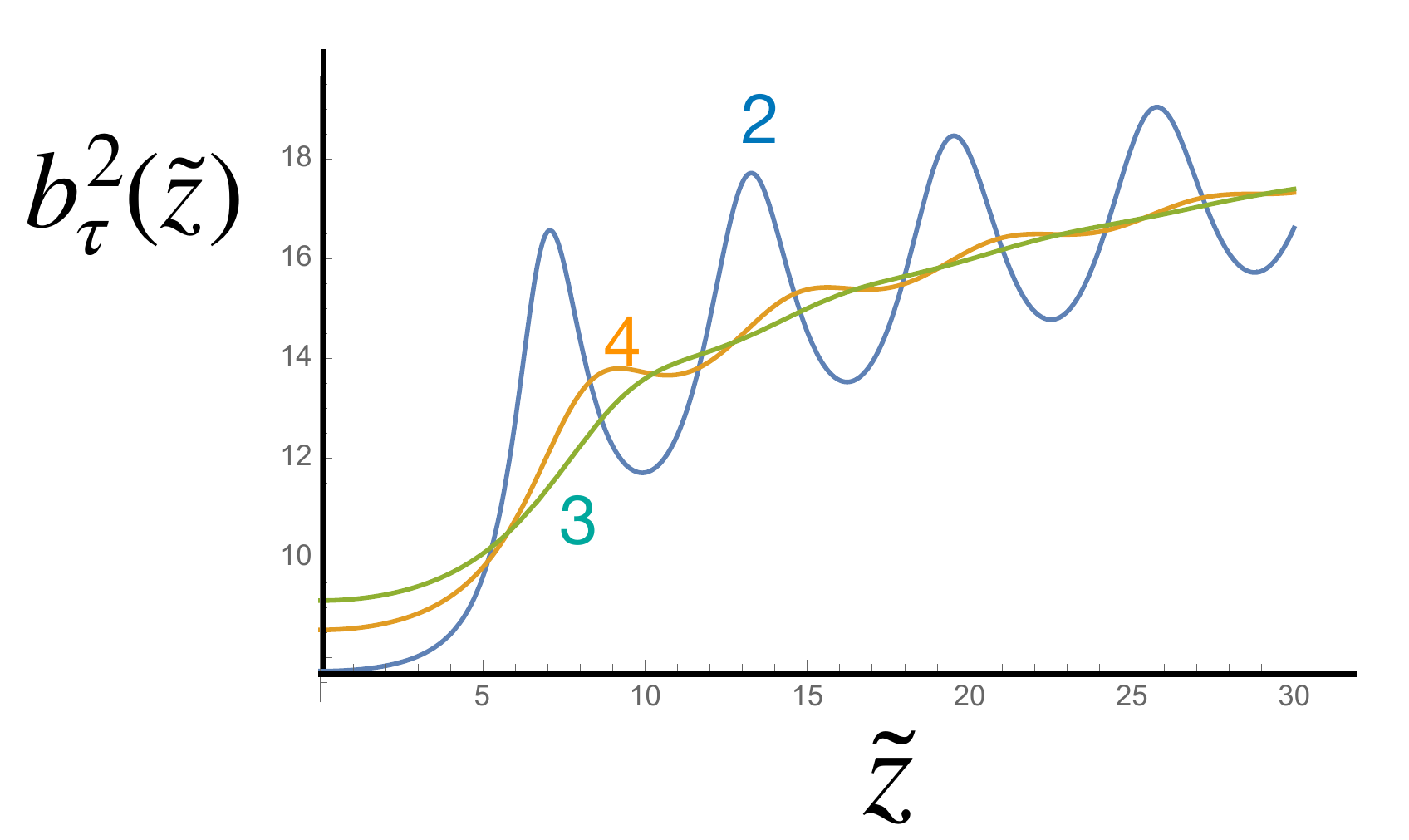}
 \caption{$b^2_\t(\tilde z)$ in units of GeV$^{-2}$. The numbers refer to the value of $\t$,  the number of constituents in the Fock state.}\label{bsqz}\end{figure}

We examine how the transverse extent depends on $\tz$  
 by defining an expectation value
{ $ b^2_\t(\tz)\equiv {\int d^2b \left\vert\psi_\t(\tz,b)\right\vert^2 b^2\over  \int d^2b \left\vert\psi_\t(\tz,b)\right\vert^2}. $}
The values  shown in Fig.~\ref{bsqz}  generally increase with increasing $\tz$,  in contrast with  intuition based on rotational invariance.
  \\

 The unorthodox behavior shown in Fig.~\ref{bsqz} motivates us to  define the  average  value of $\bfb^2$ as a function of $x$: 
 $ b^2(x)\equiv \int d^2{b} b^2 \left \vert  \psi_{\it eff}(x, \mbf{b} ) \right\vert^2= {4f\over (1-x)^2}.$  This quantity is independent  of the value of $\t$ and  ranges from about 1.1 fm$^2$ at $x=0$ to 0.23 fm$^2$ at $x=1$.  This {  decreasing} behavior arises from  the vanishing  of $f(x)$ as $x$ approaches 1.  Indeed  $\lim_{x\to1}f(x)\,{\rm GeV}^2=1.27454 (1-x)^2+0.416245 (1-x)^3+\cdots$.
The mean-square  transverse size decreases with increasing $x$.  Similarly,  the mean-square transverse momentum $k^2(x)= 1/b^2(x) 
$ increases with increasing $x$.  
This behavior is completely opposite to that obtained from the simpler form of \eq{wf}, as well as that of many models of GPDs. \\

 \medskip\noindent {\it Summary and Outlook} 
 
 A longitudinal spatial variable $\tz$ has been introduced in \eq{tildez}, thus allowing a representation of light-front wave functions in terms of all  three frame-independent spatial coordinate variables. 
  Both  the valence model of \eq{wf} and the universal light-front model
 which incorporates Regge behavior \eq{LFWFb} provide a light front coordinate-space density, \eq{lfd}, that has a long tail in the longitudinal separation between the struck constituent and the spectators.  This allows   the absorption-emission separation distance, $\xm$  occurring in deep inelastic scattering  to be very large. The result \eq{gfinal}  shows how given regions of $\xm$ contribute to the quark distribution at each value of  Bjorken $x$.  \\


 \medskip
{\it Acknowledgments}
 This   work has been partially supported by 
U.S. D. O. E.  Grant No. DE-FG02-97ER-41014 and by 
 U.S. D. O. E.  contract. No.  DE--AC02--76SF00515.  SLAC-PUB-17497. {  We thank T.~Liu, R.~Suffian, A.~Radyushkin, D.~Richards and J.~Qiu for  useful discussions.}


\end{document}